\documentclass{article}

\begin{document}

\centerline{\Large \bf
Summary of Session D1(ii), String Theory and Supergravity}  

\centerline{Donald Marolf}   

\centerline{Physics Department, Syracuse University\\
            Syracuse, NY 13244-1130, USA}  


\begin{abstract}
The talks presented in the string theory and supergravity session 
of the GR16 conference in Durban, South Africa are described below for
the proceedings.
\end{abstract}

The strings and supergravity session featured a small but varied collection
of talks ranging from studies of exact solutions and solitons to supersymmetry, 
cosmology, and talks related to the gauge-theory/gravity dualities of\cite{malda,ISMY}.
Given the breadth of topics and the rather liberal amount of space made available 
in the proceedings, it seemed best to allow each speaker to describe their talks
in some length.  What follows is therefore a description of each talk in the order that they
were presented.  Each contribution was written by the speaker and only
slightly edited by myself as session chair.  For full details of the works, please
refer to the references provided below.

\vskip \baselineskip
\begin{center}
{\bf Reduced $D=10$ ${\cal N}=4$ Yang-Mills Theories}

Matthias Staudacher
\end{center}

Staudacher discussed various aspects of the dimensional reductions of the
maximally supersymmetric gauge theory, namely ${\cal N}=1$ in $D=10$
to lower dimensions, and in particular the cases
$D=10 \rightarrow d=0,1,2,4$. These reductions are relevant since
supersymmetry survives the reduction process, allowing one to obtain
a number of exact and analytic results. This is particularly
important since there exist manifold, largely conjectural relationships to
string theory, supermembrane quantization and supergravity models.

The first half of the talk focused on the reductions to
$d=0,1,2$. All three cases have been used in different proposals
for non-perturbative definitions of string theory and M-theory,
going, respectively, by the names IKKT model, 
deWit-Hoppe-Nicolai or BFSS model, DVV model or matrix string theory.
A report was given on a number of recent results concerning the d=0 reduction\cite{Krauth:2000bv,Staudacher:2000gx}.  This reduction is of interest
in its own right and is also relevant to the bound state problem
of the $d=1$ reduction.  This work demonstrates that the current
techniques for counting the number of ground states are not yet
consistent and in fact quite incomplete. It is also relevant
to the issue of computing exact partition functions in the
$d=2$ reduction.

The second half of the talk discussed an ongoing investigation
of Maldacena-Wilson loops in the $d=4$ gauge field theory: Using the
AdS/CFT correspondence between classical supergravity and the
strong coupling limit of the field theory, a number of exact results
for these loops have been proposed in the literature. An important
problem consists in relating these results to weakly coupled
gauge theory. The first-ever two-loop
perturbative calculation\cite{Plefka:2001bu} was discussed.
The first main result is that the Maldacena-Wilson loop operator is
completely two-loop finite, suggesting finiteness to all orders.
The second chief result is that vertex diagrams contribute to
the two-loop static potential. Previous lower order calculations in
the literature found only (trivial) ladder diagrams to contribute.
This indicates that, if AdS/CFT is correct, it indeed solves,
at strong coupling, 't Hooft's longstanding planar diagram 
summation problem for this field theory.

\vskip \baselineskip
\begin{center}
{\bf Rotating black holes in higher dimensions}

Roberto Emparan
\end{center}

There are several motivations for studying General Relativity in
dimensions higher than 4. In addition to
allowing rich dynamics and new qualitative behavior, higher dimensions
are required by most unification
schemes, such as string theory. Moreover, in scenarios with large extra
dimensions and a low fundamental scale there is the possibility
that black holes will be produced and detected in future colliders\cite{bhlhc}. Such black holes will generically be rotating.

Solutions describing neutral, rotating black holes in higher
dimensions were found in\cite{mp}. For $D\geq 6$ they present new
qualitative features which, perhaps surprisingly, have so far attracted little
attention. This is the subject of this contribution, which is
based on work in progress with Rob Myers.

In $D\geq 5$ dimensions, rotation can take place in more than one
plane. However, the talk considered the case where the black hole spins in a single
plane, with rotation parameter $a\propto J/M$. 
As is well known, the
amount of rotation that a 4D Kerr black hole can support is limited: if
the bound $M\geq a$ is violated, a naked singularity results. As shown
in\cite{mp}, a similar bound appears for $D=5$. However, for $D\geq 6$
the horizon is present for arbitrary values of $a$. There is no
extremal limit and no bound on the spin.
Hence, these ultra-spinning black holes are distinctive of higher
dimensions.

What is the shape of such an ultra-spinning black hole? An analysis of
the proper size of the horizon reveals that the black hole is highly
flattened along the directions parallel to the plane of rotation---a
`pancaked' horizon. Moreover, if the rotation parameter is sent
to infinity (with the mass per unit area kept finite), the geometry
that results is that of a black membrane. The latter is known to be
classically unstable\cite{gl}, so it is natural to ask whether the
highly pancaked, ultra-spinning black holes will not become unstable
before reaching this limit, i.e., already at finite values of $a$.

The local stability of 4D black holes has been established in classic
work\cite{chandra}. Global stability has been addressed using the area
theorems (and their thermodynamic interpretation), which state that the
sum of the areas of the future horizons can not decrease. As for black
branes, both local and global arguments lead to the conclusion that
they are unstable\cite{gl}.

Both approaches should be applied to 
ultra-spinning black holes. 
While the local analysis of linear perturbations is
still in progress, global thermodynamics points clearly
to an instability of black holes in $D\geq 6$ for sufficiently large
values of $a/M$.  To see this, the possible decay modes
of the black hole were identified and the area of the 
final products were compared to
that of the initial black hole for the same mass and spin. 
Several decay modes were studied, such as emission of waves (modeled by a
gas of null particles, or using the radiation formulas for
near-Newtonian sources), or the black hole breaking apart into several
smaller black holes (a classically forbidden process). In all cases the
instability sets in at $a/M\approx$ a few. Consistently, this happens
only for $D\geq 6$; for $D=4,5$ this approach predicts no instability.

\vskip \baselineskip
\begin{center}
{\bf Black hole entropy calculations based on symmetries}

Jacek Wi\'sniewski, with Olaf Dreyer and Amit Ghosh
\end{center}

A microscopic derivation of black hole entropy is one of the
greatest challenges to candidate quantum theories of gravity. As
an alternative to the existing quantized models, a set of very
attractive ideas was recently suggested by Andrew Strominger\cite{strom}
and Stephen Carlip\cite{carlip}. Their
symmetry based approaches are very general and mostly classical.
The starting point is an observation that symmetry generating
vector fields of black hole space-time form a Diff($S^1$) algebra
which, on the level of the algebra of conserved charges associated
with these vector fields, becomes a centrally extended Virasoro
algebra\cite{bh}. The black hole space-time turns out to be a
representative state of the algebra with a fixed conformal weight.
Then, solely using representation theory, one can count the
degeneracy of such a state and this gives the entropy of the black
hole : $S=2\pi\sqrt{cH_0/6}$. Here, $c$ is the central extension
and $H_0$ is the conformal weight of the black hole or, more
precisely, the eigenvalue of the Hamiltonian (same as the
zero-mode of the Virasoro algebra) for the black hole space-time
as the eigenstate.

These calculations, however, face some conceptual and technical
problems. Strominger's calculations are based on {\it asymptotic}
symmetries. It is not apparent how these symmetries capture the
essence of the black hole space-time. In fact, the results are
equally applicable to a star having similar asymptotic behavior.
Subsequently, Carlip improved on this idea by making the symmetry
analysis in the near-horizon region. Conceptually this approach is
much more satisfactory in that the black hole geometry is now at
the forefront. Ref.\cite{carlip} also, however, faces some technical
problems which corrected in\cite{dgw} and some interesting
results emerge :

a) The Lie brackets of the vector fields form a Diff($S^1$)
algebra both on and near the horizon. b) The vector fields do not
admit a well-defined limit to the horizon 
(horizon penetrating coordinates were used to make this explicit). 
c) It is essential
that the entire calculation of the Poisson bracket of charges
(i.e., Hamiltonians of the vector fields) is performed at a
distance $\epsilon$ away from the horizon and the limit
$\epsilon\to 0$ is taken at the end. The lack of limit of the
vector fields forbids a clear interpretation of the symmetries in
the classical gravity. Presumably, this feature is an indication
that the true origin of the central charge is quantum mechanical
where $\epsilon$ is to be regarded as a regulator.

There is in fact a whole one-parameter family of such
vector fields forming Diff($S^1$), as above. As a result both $c$
and $H_0$ are modified, but in such a way that the entropy is
always reproduced (up to a multiplicative factor of $\sqrt 2$).
Only for a specific choice of this parameter, motivated by some
arguments in the Euclidean signature, one can shift $H_{0}$ by
some `ground state' value to get rid of $\sqrt 2$. However, the
meaning of this choice is unclear.

 Finally, one can perform a symmetry analysis completely on the
horizon within the framework of {\em isolated horizons} (see other
talks). The framework is naturally suited to address the question
of symmetries of black holes in equilibrium (very weak assumptions
are made). It turns out that no central extension appears in this
case. Summarizing, the current viewpoint of the authors is that
probably there is some truth in the symmetry based approaches, but
it is hard to avoid details of quantum theory, especially in order
to understand the origin of the central charge. This could be the
case because one is attempting to give an essentially classical
argument for a phenomenon that is inherently quantum mechanical.

\vskip \baselineskip
\begin{center}
{\bf Exact Super Black Hole Solutions}

R.B. Mann,
with J. Kamnitzer and
M.E. Knutt 
\end{center}

Mann reported on a project
which considered the problem of finding exact solutions to
supergravity coupled to matter. Very few exact classical solutions to
supergravity theories are known that have non-trivial fermionic content. 
A superparticle (which, if massive, is a D0-brane), a
cosmological constant and a super-Liouville field were included as 
matter sources,
each minimally coupled to $(1+1)$-dimensional supergravity.

One of the pitfalls of finding exact solutions is in ensuring that they
cannot be reduced by infinitesimal local supersymmetry transformations to
purely bosonic solutions. Working in
superspace offers a straightforward means of avoiding this difficulty, since
a superspace supergravity solution -- one which satisfies the constraints --
has nonzero torsion beyond that of flat superspace.  The torsion is a
supercovariant quantity, and as such its value remains unchanged under a
gauge transformation. Hence any exact superspace solution with non-zero
torsion must necessarily be non-trivial in this sense.

Mann reported on several exact solutions obtained for each of the supermatter
sources mentioned above. The exact compensator superfield that describes the
supergravity can be used to construct models of two-dimensional
supersymmetric black holes with non-trivial curvature.  To our knowledge
these are the first solutions found using this technique. \ The
superparticle and cosmological solutions had locally constant supercurvature\cite{RM1}, but 
the super-Liouville solution had locally non-constant curvature\cite{RM2}.  
In the latter case possibility that a gravitini
condensate formed was considered 
and examined the implications for the resultant spacetime
structure. All such condensate solutions were found to have a condensate
and/or naked curvature singularity.

\vskip \baselineskip
\begin{center}
{\bf Inverse dualisation and non-local dualities
between gravity and supergravity theories}

Dmitri Gal'tsov,
with Chiang-Mei Chen and Sergei A. Sharakin
\end{center}

Gal'tsov's talk was based on the preprint hep-th/0109151 which finds classical
dualities of a new type  between Einstein vacuum gravity in certain
dimensions and ten and eleven-dimensional supergravities. The main idea is
that Kaluza-Klein two-forms arising in toroidal compactification of vacuum
gravity can be dualized in dimensions $D\geq 5$ to higher rank antisymmetric
forms and these forms may be identified as matter fields belonging to
bosonic sectors of supergravities.  While it is perhaps
not surprising that the Maxwell equations and the Bianchi
identities for the KK fields translate into similar equations
for dual higher rank forms, a non-trivial test is whether the
dilatonic exponents in the reduced actions are the same.
Several cases of such dualities are described. The most interesting
is the correspondence between $2+3+6$ dimensional reduction of
the eleven-dimensional supergravity and eight-dimensional
Einstein gravity with two commuting Killing vectors. A related
duality holds between both (suitably compactified) IIA and IIB
ten-dimensional supergravities and eight-dimensional Einstein
gravity with three commuting Killing vectors. Another case is
the correspondence between the ten-dimensional Einstein gravity
and a suitably compactified IIB theory. It is worth noting that
all dualities of this sort are non-local in the sense that
variables of one theory are related to  variables of the dual
theory not algebraically, but via solving differential
equations.

A remarkable fact is that the $11D$-supergravity/$8D$-gravity
duality holds not only in the bosonic sector, but also extends
to Killing spinor equations exhibiting unbroken supersymmetries
of the $11D$ theory. Namely, the existence of Killing spinors
in the supergravity framework is equivalent to the existence of
covariantly constant spinors in the dual Einstein gravity. It
would be interesting to check whether this correspondence found
at the linearized level extends non-linearly, i.e. holds for
suitably supersymmetrized $8D$ gravity. A more challenging
question is whether classical dualities found here have
something to do with quantum theories. Although an
answer was not presented in the talk, the results concerning the
ten-dimensional supergravities look promising in this direction.

\vskip \baselineskip
\begin{center}
{\bf New supersymmetry algebra on 
gravitational interaction 
of Nambu-Goldstone fermion}

Motomu Tsuda,
with Kazunari Shima 
\end{center}

A supersymmetric composite unified model for spacetime 
and matter, superon-graviton model (SGM) based upon SO(10) 
super-Poincar\'e algebra, is proposed 
in the papers\cite{ks1,ks2}. 
In SGM, the fundamental entities of nature are the graviton 
with spin-2 and a quintet of superons with spin-1/2. 
The fundamental action which is the analogue 
of Einstein-Hilbert (E-H) action of general relativity (GR) 
describes the gravitational interaction 
of the spin 1/2 N-G fermions in Volkov-Akulov (V-A) model\cite{va} 
of a nonlinear realization of supersymmetry (NL SUSY) 
regarded as the fundamental objects (superon-quintet) for matter. 

Tsuda's talk performed the similar geometrical arguments 
to GR in the SGM spacetime, where the tangent Minkowski spacetime 
is specified by the coset space SL(2,C) coordinates 
(corresponding to N-G fermion) of NL SUSY of V-A model\cite{va} 
in addition to the ordinary Lorentz SO(3,1) coordinates, 
and discussed the structure of the fundamental SGM action\cite{ks2,st1,st2}. 
The overall factor of SGM action is fixed to ${-c^3 \over 16{\pi}G}$, 
which reproduces E-H action of GR in the absence of superons (matter). 
Also in the Riemann-flat space-time, i.e. the vierbein $e{_a}^{\mu}(x) 
\rightarrow \delta{_a}^{\mu}$, 
it reproduce V-A action of NL SUSY\cite{va} with 
${{\kappa}^{-1}}_{V-A} = {c^3 \over 16{\pi}G}{\Lambda}$ 
in the first order derivative terms of the superon. 
Therefore our model (SGM) predicts a (small) non-zero cosmological 
constant, provided ${\kappa}_{V-A} \sim O(1)$, 
and possesses two mass scales. 
Furthermore it fixes the coupling constant of superon (N-G fermion) 
with the vacuum to $({c^3 \over 16{\pi}G}{\Lambda})^{1 \over 2}$ 
(from the low energy theorem viewpoint), 
which may be relevant to the birth 
(of the matter and Riemann space-time) of the universe. 
The (spacetime) symmetry of 
our SGM action was also demonstrated.  In particular, 
the commutators 
of the new NL SUSY transformations on gravitational 
interaction of N-G fermion with spin-1/2\cite{ks2,st1} 
and -3/2\cite{st1,st3} form a closed algebra, which reveals 
N-G (NL SUSY) nature of fermions and the invariances 
at least under a generalized general coordinate 
and a generalized local Lorentz transformations. 
In order to linearize the SGM action, the linearization 
of $N = 2$ V-A model which is now under investigation 
is extremely important from the physical point of view, 
for it gives a new mechanism generating a (U(1)) gauge field 
of the linearized (effective) theory\cite{ks3}.

\vskip \baselineskip
\begin{center}
{\bf Quantum Cosmology from 
D-Branes}

P. Vargas 
Moniz,
with A. Yu. Kamenshchik
\end{center}

Recent developments in string theory suggest that, in a Planck length 
regime, the quantum fluctuations are very large so that string coupling increases and 
consequently the string degrees of freedom would not be the relevant ones.
Instead, solitonic degrees of freedom such as D-p-branes would become more 
important. Hence, what would be the effect of those new physical 
degrees of freedom on, say, the the very early universe and in particular 
from a quantum mechanical point of view?

In this work, one has initiated an investigation on D-$p$-brane induced
quantum  cosmology. It could be pointed that it may not be justified to
quantize  an effective theory (arising from a fundamental quantum theory).
However, in so far  as new fundamental fields and effects arise from the
fundamental theory, a quantization of the effective action could capture
significant and relevant novel features.

The starting point  is the result (obtained by Duff, Khai and Lu)
that the natural metric that couples to a $p$-brane is the Einstein metric multiplied 
by the dilaton. Employing an (adequately) modified Brans-Dicke action with a 
deformation parameter $p$-dependent, different quantum cosmological scenarios
were  analyzed. In particular, several early universe scenarios were
identified  similar to quantum Pre-Big-Bang and Universe-anti-Universe
creation.  The possible quantum mechanical transition amplitudes were also
studied with a view towards determining the effect of quantum 
cosmological solitons in the very early Universe. 
Finally, it was found that the solutions of Wheeler-DeWitt equation 
allowed for sub-class with 
$N=2$ SUSY. Other  consequences regarding  
"stringy" cosmology were investigated, namely possible duality
transformation in the effective action
and their relation to Pre-Post-Big-Bang scenarios. 
Possible developments of this work include 
considering  D-brane realistic actions and SUSY extensions.


\vskip \baselineskip
\begin{center}
{\bf Polarization of the D0 ground state in quantum mechanics and 
supergravity}

Donald Marolf,
with Pedro Silva
\end{center}

Marolf's talk addressed a
quantum version of the dielectric
effect described by Myers in\cite{mye1}.  In\cite{mye1},
the application of a Ramond-Ramond background field
to a D0-brane system induces a classical dielectric effect
and causes the D0-branes to deform into a non-commutative D2-brane.
In contrast,\cite{pedro} places D0-branes in the background 
generated by a stack of D4-branes.  While
no classical dielectric effect results, the four-branes modify
the potential that shapes the non-abelian character of the
quantum D0 bound state.  As a result, the bound
state is deformed, or polarized.  

Two aspects of the
deformation were studied and compared with the corresponding
supergravity system.  Fundamental to this comparison is
the connection described by Polchinski\cite{pol1} relating the size of the matrix
theory bound state to the size of the bubble of space that is well-described
by classical supergravity in the near D0-brane spacetime.
The near D0-brane spacetime is obtained by taking a particular limit in which
open strings decouple from closed strings.  The result is a
ten-dimensional spacetime with
small curvature and small string coupling when one is reasonably
close (though not too close) to the D0-branes.  However, 
beyond some critical distance $r_c$ the curvature reaches the string scale.
As a result, the system beyond $r_c$
is not adequately described by the massless fields of classical supergravity.
The goal was thus to compare deformations of the non-abelian D0-brane
bound state with the deformations of this bubble of `normal' space.

While the detailed effects were beyond the scope of the work presented, 
 the deformations of the quantum mechanics ground state and the supergravity bubble
were shown to have corresponding scaling properties.  This supports the idea that
the gravity/gauge theory duality associated with D0-branes can be
extended to include couplings to nontrivial backgrounds such as those
discussed in\cite{tvr1,tvr2,mye1}.  A part of this was the analysis  in
an appendix of infrared issues associated with 't Hooft scaling in 0+1
dimensions.  This in turn strengthens the argument that Polchinski's
upper bound\cite{pol1} on the size of the D0-brane bound state in fact
gives the full scaling with $N$.  Corresponding arguments can in fact be made
for all Dp/D(p+4)-systems for $p\le 2$.

\end{document}